\documentclass{article}
\usepackage{graphicx}
\usepackage{amsmath}
\usepackage{amsfonts}
\usepackage{amssymb}
\usepackage{hyperref}
\usepackage{enumerate}
\usepackage{color}
\newcommand{\etc}{\textit{etc.}}
\newcommand{\fminus}{\ensuremath{
\displaystyle
\frac{\alpha E^2}{16\pi\hbar}
}}
\newcommand{\fplus}{\ensuremath{
\displaystyle
\frac{\alpha^2 E^4 m L^2}{256\pi\hbar^3} +
\frac{\hbar}{4\pi m L^2}
}}

\author{L.A.\,Melnikovsky\thanks{E-mail: leva@kapitza.ras.ru}\\
\\
\textit{P.L.\,Kapitza Institute for Physical Problems}\\
\textit{Russian Academy of Sciences, Moscow, Russia}}
\title{Roton Parametric Resonance}
\date{}
\begin{document}
\maketitle

\begin{abstract}
Parametric excitation of rotons by oscillating electric field
exhibits a narrow resonance at the roton minimum frequency.
The resonance region width is in good agreement with experimental
results on the microwave absorption in superfluid helium.

\end{abstract}

\section{Introduction}
Recent experiments with a whispering gallery resonator submerged in
superfluid $^4$He produce striking result. The electromagnetic damping
exhibits an ultra-narrow resonance \cite{ryba1,ryba2,ryba3}. The resonance frequency
depends on the temperature and tracks precisely the roton energy gap.

Such results were truly unexpected:
\begin{enumerate}[(A)]
\item \label{narrow} Energy spectrum of helium is continuous, excitations exist
both below (phonons) and above (phonons and rotons) the roton energy
gap. The ultra-narrow absorption line can not be explained with plain
single-particle density of states peculiarities. To the best of our
knowledge, no realistic explanation has been suggested for this.
\item Helium is believed to be perfectly neutral and has relatively
low susceptibility ($\epsilon=1.05$). It is therefore not obvious how
electromagnetic oscillations interact with superfluid excitations.
Conceivable scenarios include polarization of liquid by motion
\cite{pash,gravel1,gravel2} or vulgar acoustic emission by mechanic vibrations
of the resonator walls due to electrostriction.
\item \label{alienmomentum}
Creation of a roton requires significant momentum.
Photon momentum is much smaller and can not account for the momentum
conservation. This process was therefore thought to be possible
\cite{ryba1} only in the immediate vicinity of the wall which breaks translation invariance.
\end{enumerate}
These difficulties motivated present research.

To deal with the problem \eqref{narrow} we assume that observed effect
is essentially collective. Particularly, rotons are bosons, emission of
new ones is stimulated by those already present. It is important, that
new rotons are emitted into the same quantum state. The avalanche
occurs if the gain (number of rotons emitted per unit time) is bigger
than the losses (number of rotons leaving active volume per unit time).
The latter is proportional to the roton velocity and is arbitrary small
near the minimum energy.

Similar phenomenon is realized in lasers, where photons are usually kept
in a resonant cavity to stimulate emission of subsequent photons.
One does not need a resonator for rotons, they remain in the active
volume due to their slowness.

%conventional coupling between electric field and medium density variations,
Without excessive speculations we stick to the traditional \cite{LL8}
expression $\epsilon \mathbf{E}^2/(8\pi)$ for the energy density in the electric field.
This term amounts to the field-dependent change in the
energy of elementary excitation. Particularly for the rotons \cite{andreev}
\begin{equation}
\label{roten}
\varepsilon=\Delta+\frac{(p-p_0)^2}{2m}+\alpha_{ij} \frac{E^i E^j}{2},
\end{equation}
where
$\Delta=8.7\,\text{K}$, 
$p_0=2\cdot 10^{-19}\,\text{g}\cdot\text{cm}/\text{s}$, and
$m=0.16\,m_\text{He}$ are the roton spectrum parameters. Precise value of the the factor
$\alpha$ (roton polarizability) is unknown\footnote{Experimentally, the resonance line is split in two
when a static electric field is applied. We may attribute the second order
term in the frequency displacement of a resonance component to the polarization of rotons. This
gives \cite{rybapriv} $\alpha=2.1 \cdot 10^{-25}\,\text{cm}^3$.} but
can be estimated \cite{andreev} as following
\begin{equation}
\alpha \sim \frac{(\epsilon-1)^2 p_0^2}{4 \pi \rho \Delta} \sim 
4.5 \cdot 10^{-26}\,\text{cm}^3,
\end{equation}
where $\rho=0.145\,\text{g}/\text{cm}^3$ is helium density.

Suppose the electric field oscillates with the frequency $2\pi f$. The
roton energy \eqref{roten} ``oscillates'' twice as fast because the
last term is quadratic with respect to the electric field. This means
that the electric field parametrically pumps the rotons and their
population grows exponentially if the pumping frequency $4\pi f$ is
close to the double roton frequency $2\omega=2\varepsilon/\hbar$. Note
also, that the problem of seemingly unconserved momentum \eqref{alienmomentum} is
irrelevant for the parametric roton excitation: two
rotons are created simultaneously.
%with opposite momenta.

\section{Resonance width}
\begin{figure}
\begin{center}
\def\svgwidth{110mm}
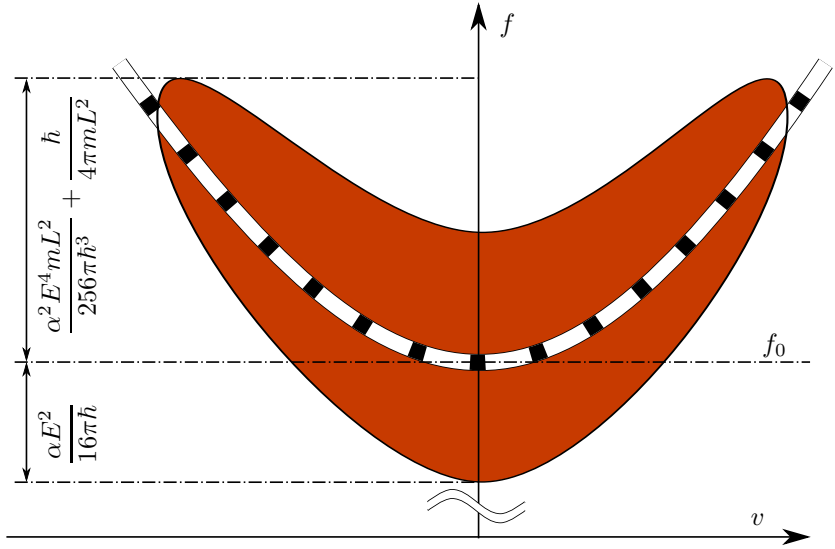
\end{center}
\caption{Grin-like area corresponds to the parametric instability \eqref{instability} region (not to scale).
Roton dispersion $2\pi\hbar f=mv^2/2$ is plotted with dashed line. 
}
\label{grin}
\end{figure}

As explained above, a coherent roton state with large occupation
number is formed in a parametric resonance. This enables a
classical \cite{LL1} treatment of the roton field.
In a mechanical system with the time-dependent intrinsic frequency
$\omega(t)=\omega+\gamma \cos 4\pi f t$,
the parametric resonance occurs if 
\begin{equation}
\label{condition}
\left| 4\pi f - 2\omega \right| < \sqrt{\gamma^2-4\lambda^2}.
\end{equation}
The damping decrement $\lambda$ for rotons is determined
by the time they spend in the active volume. If the pumping
field is confined in a volume with linear dimension $L$, then
\begin{equation}
\label{damping}
\lambda \sim v/L,
\end{equation}
where $v=(p-p_0)/m$ is the roton velocity.
Assuming $E(t)=E\cos 2\pi ft$ we get
\begin{equation}
\label{gain}
\gamma \sim \frac{\alpha E^2}{4\hbar}.
\end{equation}
Substituting \eqref{roten}, \eqref{damping}, and \eqref{gain} into \eqref{condition}
we obtain
\begin{equation}
\label{instability}
\left| 4\pi f 
-\frac{2\Delta}{\hbar}
-\frac{mv^2}{\hbar}
\right|
\lesssim
\sqrt{\left(\frac{\alpha E^2}{4\hbar}\right)^2- \frac{4 v^2}{L^2}}.
\end{equation}
This inequality can be satisfied (at varying velocities, see Fig.\ref{grin}) if
\begin{equation}
\label{linewidth}
-\frac{\alpha E^2}{16\pi\hbar}
\quad \lesssim \quad 
f-f_0
\quad \lesssim \quad 
\frac{\alpha^2 E^4 m L^2}{256\pi\hbar^3} + \frac{\hbar}{4\pi m L^2},
\end{equation}
where $f_0=\Delta/(2\pi\hbar)\sim 180\,\text{GHz}$.

\section{Discussion}
In the whispering gallery resonator experiments the absorption line
width at low temperature could not be reliably measured. Instrumental
frequency resolution was on the order of $50\,\text{kHz}$. At higher
temperature the line width increases and reaches $350\,\text{kHz}$.

In these experiments the oscillating electric field reaches as much as
$E\sim 200\,\text{V}/\text{mm}$ on the resonator wall and decays
exponentially outward in helium. The length scale is on the order of
magnitude of the microwave length  $L\sim 1\,\text{mm}$. Substituting
these numbers in \eqref{linewidth} we get
\begin{equation*}
-40\,\text{Hz}
\quad \lesssim \quad 
f-f_0
\quad \lesssim \quad 
45\,\text{kHz}.
\end{equation*}
This gives the full resonance region width of about 45\,kHz at low
temperature. Exact spectral line shape (not Lorentzian) is determined
by the nonlinear terms.

Several factors should contribute to the parametric instability widening at
finite temperature. Roton scattering explicitly increases the roton
decay and the natural roton linewidth (neglected above). It also
reduces their active volume escape rate, effectively retarding them.

We emphasize, that a coherent roton state (a $\mathbb{ROTON}$) is formed
by the parametric excitation. In this sense the electromagnetic
resonator and the superfluid around it work as a ``laser of rotons'' (raser).

This is the roton slowness that is responsible for the narrow resonance.
Another stationary point of the superfluid dispersion curve corresponds
to maxons, the same effect is expected at the respective frequency
$f_\text{M}\sim 300\,\text{GHz}$. Present analysis is not specific to
superfluid, other quasiparticles (magnons in magnetics, photons in
metamaterials, \etc) with zero velocity will exhibit similar behavior. 
Note, that possibility of parametric excitation of magnons \cite{suhl}
and photons \cite{akhmanov} is well established.

Higher order parametric resonances are possible at fractional
frequencies. It would be interesting to perform accurate measurement of
the resonator quality factor near $f_0/2$, $f_0/3$, \etc\  Another
contribution is the result of linear coupling to the electric field
\cite{quadrupole} due to the roton quadrupole moment. This will lead to
the double frequency resonance at $2f_0$.

\section{Acknowledgements}

I thank A.F.\,Andreev, V.I.\,Marchenko, V.P.\,Mineev, and A.S.\,Rybalko for fruitful discussions.
This work was supported by RF president program NSh-4889.2012.2.

\end{document}